\documentclass{elsart}

\usepackage{amsmath,amssymb,bm}

\newcommand{\tr}{\mathop{\mathrm{Tr}}}
\newcommand{\re}{\mathop{\mathrm{Re}}}
\newcommand{\bacol}{\setlength{\arraycolsep}{0pt}}
\newcommand{\python}{\textsc{Python}}
\newcommand{\tA}{\tilde{A}}

\journal{Journal of Computational Physics}

\begin{document}

\begin{frontmatter}


  
\title{Automatically generating Feynman rules for improved lattice
    field theories}


\author[edinburgh]{A. Hart\corauthref{cor}},
\corauth[cor]{Corresponding author.}
\author[cambridge,regina]{G.M. von Hippel},
\author[cambridge]{R.R. Horgan\corauthref{cor}},
\author[cambridge]{L.C. Storoni}.

\address[edinburgh]{School of Physics, University of
Edinburgh, King's Buildings, \\ Edinburgh EH9 3JZ, U.K.}

\address[cambridge]{DAMTP, CMS, University of Cambridge, Wilberforce Road,
\\ Cambridge CB3 0WA, U.K.}

\address[regina]{Department of Physics, University of Regina, Regina,
  SK, S4S 0A2, Canada.}

\begin{abstract}
  
  Deriving the Feynman rules for lattice perturbation theory from
  actions and operators is complicated, especially when improvement
  terms are present. This physically important task is, however,
  suitable for automation.  We describe a flexible algorithm for
  generating Feynman rules for a wide range of lattice field theories
  including gluons, relativistic fermions and heavy quarks. We also
  present an efficient implementation of this in a freely available,
  multi-platform programming language (\python), optimised to deal
  with a wide class of lattice field theories.

\end{abstract}

\end{frontmatter}


\section{Introduction}

Non--abelian quantum field theories such as QCD are believed to
explain much of particle physics, at least at energy scales probed by
current particle accelerators. Perturbative expansions of the theory
do not, however, converge at hadronic energy scales. That, and the
belief that non-perturbative physics may also contribute to certain
states, makes the lattice regularisation of quantum field theories
extremely important. Inherently non-perturbative calculations can then
be carried out using Monte Carlo simulations.  

Dividing space and time into a grid with lattice spacing $a$, however,
excludes ultraviolet modes with momenta of $\pi/a$ or higher. A
renormalisation programme is therefore required to connect lattice
measurements to their continuum counterparts. Such renormalisation
factors are particularly important for QCD matrix elements and fixing
the couplings and masses present in the Lagrangian. Renormalisation is
also needed to determine the strong coupling $\alpha_s$ and to relate
the lattice regularisation scale $\Lambda_{\mathrm{lat}}$ to the more
familiar $\Lambda_{\mathrm{QCD}}$. It is also used to ``improve'' the
lattice actions in an attempt to reduce the discretisation errors at
given lattice spacing.

In a limited number of cases the renormalisation constants can be
determined using non-perturbative techniques. Results at finite
lattice spacing, however, can depend upon the method used (e.g.
\cite{Bhattacharya:2000pn}),
and non-perturbative methods do not cope well with mixing of operators
under renormalisation. For these reasons there is a strong interest
in lattice perturbation theory.

Given that perturbation theory fails in low energy QCD, we may ask why
it should work on the lattice. An argument for its use is given in
\cite{Lepage:1996jw}:
the renormalisation factors may be thought of as compensating for the
ultraviolet modes excluded by the lattice regulator. For typical
lattices $a \lesssim 0.1 \textrm{~fm}$, and the excluded modes have
momenta in excess of 5 GeV. At these scales the running QCD coupling
$\alpha_s$ is small enough that perturbation theory should rapidly
converge. The wide range of results recently reviewed in
\cite{Capitani:2002mp,Trottier:2003bw}
show perturbation theory can be used for a large range of lattice QCD
processes. It is an assumption that non-perturbative effects do not
contribute on these short length scales. In a few cases we can test
this directly by comparing high order perturbative calculations with
Monte Carlo simulations at a range of weak couplings
\cite{Lepage:1999qr,DiRenzo:2000ua,Horsley:2001uy,Trottier:2001vj,Hart:2004jn}).
The non-perturbative contributions to the studied quantities are very
small. Other comparisons, such as 
\cite{Bhattacharya:2000pn},
cannot distinguish non-perturbative effects from higher loop
perturbative corrections. It therefore remains that lattice
perturbation theory provides the only systematically improvable method
for determining the full range of renormalisation constants
\cite{Capitani:2002mp}.

As in the continuum, the calculation of lattice Feynman diagrams is a
two stage process. The lattice action and operators must first be
Taylor expanded to give the propagators and vertices that form the
Feynman rules (which we refer to as the ``vertex expansion'' stage).
Following this, these rules must be used to construct and evaluate
Feynman diagrams, possibly after algebraic simplification (the
``Feynman diagram evaluation'' stage).

The main obstacles in the latter task are the presence of Lorentz
symmetry violating terms at finite lattice spacing and the
complications of replacing momentum integrals by discrete sums.  The
calculations are therefore usually done using computer programs like
\textsc{Vegas}
\cite{Lepage:1977sw},
\textsc{Form}
\cite{Vermaseren:2000nd}
or other proprietary mathematical packages.

Expanding the lattice action and operators to obtain Feynman rules is
far more complicated than in the continuum. Firstly, lattice gauge
fields are elements of the Lie group rather than the algebra of the
gauge group. We must therefore expand exponentials of non--commuting
fields to obtain the Feynman rules. Secondly, modern lattice theories
contain a large number of irrelevant (in the renormalisation group
sense of the word) terms chosen to improve specific aspects of the
Monte Carlo simulation, such as the rate of approach to the continuum
or chiral limits of QCD.

There is, however, no unique prescription for these terms, and the
choice depends on that quantities we are most interested in
simulating. As a result, a large number of actions and operators are
currently in use. Although the differences may be subtle, each choice
provides a separate regularisation of QCD with its own set of
renormalisation constants and, most relevantly here, Feynman rules. At
present the complications of the expansions has meant that the
availability of renormalisation factors has lagged far behind
developments in lattice improvement. In many cases this has restricted
the physical predictions obtained from the simulations.

As a result, there is a strong need for an automated method for
deriving lattice Feynman rules in a flexible way for a range of
different theories. The generation should be rapid enough not to
constrain our choice of action, and to avoid errors we should be able
to specify the action in a compact and intuitive manner (such as using
nested link smearing prescriptions). The evaluation of the Feynman
diagrams can be computationally intensive, and may be carried out on
costly supercomputing facilities. Parsimony and software availability
dictate that the rules should be separately calculable in advance, and
rendered in a machine readable format that can be copied to any
computer for later Feynman diagram evaluation.

In this paper we describe such a method.

Automated expansion of lattice actions%
\footnote{We shall understand the term ``actions'' to include
  measurement operators from now on.}
is not a new concept, having been described for gluonic actions by
L\"{u}scher and Weisz in 1986
\cite{Luscher:1986wf}.
An implementation of this has been used in
\cite{Nobes:2001tf,Nobes:2002uu,Nobes:2003nc}.
A similar method is employed in
\cite{Alles:1992yh}.
We present here a new algorithm suited to expansion of not only
gluonic actions, but also those of complicated relativistic fermionic
actions and heavy quarks, such as in NRQCD.  As in
\cite{Luscher:1986wf},
the expansion is independent of the boundary conditions allowing, for
instance, the use of twisted boundary conditions to regulate infrared
divergences in a gauge--invariant manner
\cite{Luscher:1984xn,Luscher:1985zq}
or otherwise change the discrete momentum spectrum
\cite{deDivitiis:2004kq}.
We also describe details of an implementation of this algorithm which
we have used for calculations of the renormalised anisotropy in gauge
theories
\cite{Drummond:2002yg,Drummond:2003qu},
to study the mean link in Landau gauge for tadpole
improvement
\cite{Hart:2004jn}
and to measure the electromagnetic decays of heavy quark systems using
NRQCD
\cite{Drummond:2002kp,Gray_progress}.
The code is flexible and can be easily extended to cope with a full
range of problems, some of which we discuss in Section~\ref{sec_conc}.
We are happy to share this code with interested readers.

The structure of the paper is as follows. Our method clearly stems from
\cite{Luscher:1986wf}
and in Section~\ref{sec_algorithm} we review their theory and
notation. The algorithm itself differs markedly from
\cite{Luscher:1986wf},
not least in being able to deal with fermionic actions, and is
described in Section~\ref{sec_exp_wil}. In Section~\ref{sec_python} we
turn to implementation of the algorithm, explaining the steps taken to
ensure the code can cope with the more complicated theories.  Whilst
the notation is tilted towards our version in the \python\ programming
language, the optimisations are clearly applicable to any realisation
of the algorithm. We present our conclusions in
Section~\ref{sec_conc}. Technical details of the the data structures
employed are relegated to Appendices.

\section{The lattice}
\label{sec_algorithm}

A cubical space-time lattice $\Lambda$ in $D$ dimensions consists of
sites labelled by a vector $\bm{x} \in \Lambda$ with components that
are integer multiples of a lattice spacing~$a$, which we will set to
be one (a (bare) lattice anisotropy can be introduced through
rescaling of coupling constants in the action
\cite{Drummond:2002yg}).
The directions of the lattice axes are labelled $\mu \in
\{1,2,\ldots,D\}$. If $\bm{e}_\mu$ is a right-handed basis set
consisting of unit vectors, we define corresponding backward vectors:
$\bm{e}_{-\mu} = -\bm{e}_\mu$.

A path consisting of $l$ links starting at site $\bm{x}$ can be
specified on the lattice by an ordered set of signed integers, $s_i \in
[-D,\ldots,-1,1,\ldots,D]$:
\begin{equation}
\mathcal{L}(\bm{x},\bm{y};\bm{s}) \equiv \{ 
\bm{x},\bm{y}; \bm{s} = [s_0,s_1,\ldots,s_{l-1}] \} \;.
\label{eqn_path_def}
\end{equation}
The $j^{\, \mathrm{th}}$ point on the path is
\begin{equation}
\bm{z}_j = \left\{
\begin{array}{ll}
\bm{x} \;, & ~~j = 0 \; ,
\\
\bm{z}_{j-1} + a \bm{e}_{s_{j-1}} \; , &
~~ 0 < j \leq l \; ,
\end{array}
\right.
\label{path}
\end{equation}
and the endpoint of the path is $\bm{y} \equiv \bm{z}_l$. 

For a periodic lattice with $L_\mu$ sites in the $\mu$ direction (and
volume $V = \prod_\mu L_\mu$) the momentum vectors are
\begin{equation}
\bm{k} = \frac{2\pi}{a} \, 
\left( \frac{\bar{k}_1}{L_1}, \ldots, \frac{\bar{k}_D}{L_D} \right) \; ,
~~~~
0 \le \bar{k}_\mu < L_\mu \; , 
~~~~
\bar{k}_\mu \in \mathbb{Z} \; , 
\end{equation}
and $\sum_{\bm{k}}$ stands for sums over the integers
$\bar{k}_\mu$. The Fourier expansion of a field $\phi$ is
\begin{equation}
\tilde{\phi}(\bm{k}) = \sum_{\bm{x}} \, 
e^{ -i \bm{k} \cdot \bm{x} } \phi(\bm{x}) \;,
~~~~~~
\phi(\bm{x})
= \frac{1}{V} \sum_{\bm{k}} \,
e^{ i \bm{k} \cdot \bm{x} } \tilde{\phi}(\bm{k}) \; .
\end{equation}
Different boundary conditions (e.g. twisted 
\cite{Luscher:1986wf,Drummond:2002yg,Hart:2004jn})
change the colour factors and momentum spectrum. Since neither are
used explicitly in the vertex expansion below, the same reduced vertex
function output can be used in each case.

\subsection{Matter fields}

We now turn to the description of lattice fields. The notation follows
\cite{Luscher:1986wf}.

The gauge field associated with a link is $U_{\mu > 0}(\bm{x}) \in
SU(N)$. Let $U$ denote the full configuration of such links. The
perturbative gauge potential associated with the link is defined
through
\begin{equation}
U_{\mu>0}(\bm{x}) = \exp \left(a g A_\mu \left( 
\bm{x} + \frac{a}{2} \bm{e}_\mu \right) \right) = 
\sum_{r=0}^\infty \frac{\left( ag A_\mu(\bm{x} + \frac{a}{2} \bm{e}_\mu) 
\right)^r}{r!}
\label{eqn_link_exp}
\end{equation}
where $g$ is the bare coupling constant. The potential $A_\mu \in
\mathop{\mathrm{alg}}(SU(N))$ is associated with the midpoint of the
link. Expanding in the anti-Hermitian generators of $SU(N)$:
\begin{equation}
A_\mu = A_\mu^a\,T_a, ~~~~ 
\left[ T_a,T_b \right] = -f_{abc} T_c, ~~~~
\tr \left( T_a T_b \right) = -\half \, \delta_{ab} \;.
\end{equation}
We define $U_{-\mu}(\bm{x}) = U^\dagger_\mu(\bm{x} - a
\bm{e}_\mu)$. 

Quark fermion fields $\psi(\bm{x})$ transform according to the
representation chosen for the generators $T_a$. From now on we assume
this to be the fundamental representation (other choices will affect
the colour factors, but not the underlying expansion algorithm).

The Wilson line $L(\bm{x},\bm{y},U)$ on the lattice associated with
the path $\mathcal{L}(\bm{x},\bm{y};\bm{s})$ is a product of links
\begin{equation}
L(\bm{x},\bm{y},U) \equiv \mathcal{L} : U = 
\prod_{i=0}^{l-1} U_{s_i}(\bm{z}_i) 
= \prod_{i=0}^{l-1} \exp \left[ \mathop{\mathrm{sgn}}(s_i) a 
g A_{|s_i|} \left( \bm{z}_i+\frac{a}{2}\bm{e}_{s_i} \right) \right] \; . 
\end{equation}
As all actions and operators can be written as a sum of Wilson lines
(possibly terminated by fermion fields that are not themselves
expanded), our goal is to efficiently render $L$ as a Taylor series in
the gauge potential in momentum space:
{\bacol
\begin{eqnarray}
L(\bm{x},\bm{y} ; A) = \sum_r \frac{(ag)^r}{r!}
\sum_{\bm{k}_1,\mu_1,a_1} \ldots
\sum_{\bm{k}_r,\mu_r,a_r}
&&
\tilde{A}_{\mu_1}^{a_1}(\bm{k}_1) \ldots
\tilde{A}_{\mu_r}^{a_r}(\bm{k}_r) \times
\nonumber \\
&&
V_r(\bm{k}_1,\mu_1,a_1 ; \ldots ; \bm{k}_r,\mu_r,a_r) \; .
\label{eqn_tayl_exp}
\end{eqnarray}
}
We can write the vertex functions $V_r$ as
\begin{equation}
V_r(\bm{k}_1,\mu_1,a_1 ; \ldots ; \bm{k}_r,\mu_r,a_r) = 
C_r(a_1, \ldots , a_r) \;
Y^{\mathcal{L}}_r(\bm{k}_1,\mu_1 ; \ldots ; \bm{k}_r,\mu_r)
\end{equation}
The matrix colour factor $C_r$ plays the role of the Clebsch--Gordan
factor:
\begin{equation}
C_r(a_1, \ldots , a_r) = \prod_{i=1}^{r} T_{a_i} \; .
\end{equation}
Up to differences in the colour trace structure of the the action
(e.g. a mixed fundamental/adjoint gauge action, and discussed in
Appendix~\ref{app_patterns}), the $C_r$ are path independent. We can
therefore represent the vertex functions more efficiently by
calculating just the expansion of the reduced vertex functions,
$Y^{\mathcal{L}}_r$ (with an appropriate description of the colour
trace structure where ambiguous). The reduced vertex function can be
written as a sum of monomials
\begin{equation}
Y^{\mathcal{L}}_r(\bm{k}_1,\mu_1 ; \ldots ; \bm{k}_r,\mu_r) = 
\sum_{n=1}^{n_r} f_n \exp \frac{i}{2} \left(
\bm{k}_1 \cdot \bm{v}^n_1 + \ldots + \bm{k}_r \cdot \bm{v}^n_r \right)
\label{eqn_y}
\end{equation}
For each combination of $r$ Lorentz indices we have $n_r$ terms, each
with an amplitude $f$ and the locations $\bm{v}$ of the $r$ factors of
the gauge potential. To simplify this expression we have suppressed
the dependence of $f$, $\bm{v}^n_i$ and $n_r$ on the Lorentz
structure. To construct $Y$ for given momenta, we apply the $\bm{k}$'s
to the position vectors of all monomials with the correct Lorentz
indices.

The $\bm{v}$'s have been drawn from the locations of the midpoints of
the links in the path $\mathcal{L}$. To avoid floating point
ambiguities, it is therefore more convenient to express the components
of all $D$-vectors as integer multiples of $\frac{a}{2}$ (accounting
for the factor of $\half$ in the exponent).

\subsection{Realistic actions: the fermion sector}

We begin our discussion of realistic lattice actions with the fermion
sector.  The most general gauge- and translation-invariant action can
be written as
\begin{equation}
S_F(\psi,U) = \sum_{\bm{x}} 
\sum_{\mathcal{W}} h_{\mathcal{W}} \bar{\psi}(\bm{x})
\Gamma_{\mathcal{W}} W(\bm{x},\bm{y},U) \psi(\bm{y}) \; .
\end{equation}
It consists of Wilson lines $W$ defined by open paths
$\mathcal{W}(\bm{x},\bm{y};\bm{s})$. Associated with each path is a
coupling constant $h_{\mathcal{W}}$ and a spin matrix
$\Gamma_{\mathcal{W}}$ (which might be unity).

Using the convention that all momenta flow into the vertex, the
perturbative expansion is
\begin{eqnarray}
S_F(\psi,A) & = & \sum_r \frac{g^r}{r!} 
\sum_{\bm{k}_1,\mu_1,a_1} \ldots
\sum_{\bm{k}_r,\mu_r,a_r}
\tA^{a_1}_{\mu_1}(\bm{k}_1) \ldots 
\tA^{a_r}_{\mu_r}(\bm{k}_r) \times
\nonumber \\
& & ~~~
\sum_{\bm{p},\bm{q},b,c} \tilde{\bar{\psi}}^b(\bm{p})
V_{F,r}(\bm{p},b ; \,\bm{q},c ; 
\, \bm{k}_1,\mu_1,a_1; 
\, \ldots ; 
\, \bm{k}_r,\mu_r,a_r) \tilde{\psi}^c(\bm{q}) \;.
\nonumber\\
\end{eqnarray}
The Euclidean Feynman rule for the $r$-point
gluon--fermion--anti-fermion vertex is $-g^r V_{F,r}$, where the
symmetrised vertex is:
\begin{eqnarray}
&& V_{F,r}(\bm{p},b ; \bm{q},c ; \bm{k}_1,\mu_1,a_1; \, \ldots ; 
\bm{k}_r,\mu_r,a_r)
= 
\nonumber \\
&&
~~~~~
\frac{1}{r!} \sum_{\sigma \in \mathcal{S}_r}
\sigma \cdot C_{F,r}(b,c;a_1,\ldots,a_r) ~
\sigma \cdot Y_{F,r}(\bm{p},\bm{q} ; 
\bm{k}_1,\mu_1; \, \ldots ; \bm{k}_r,\mu_r) \; .
\label{eqn_ferm_symm}
\end{eqnarray}
$\sigma$ is an element of the permutation group of $r$ objects,
$\mathcal{S}_r$, applied to the gluonic variables and normalised by
the factor of $(r!)$. The reduced vertex $Y_{F,r} = \sum_{\mathcal{W}}
h_{\mathcal{W}} Y^{\mathcal{W}}_{F,r}$ is the sum of contributions
from paths $\mathcal{W}$.

For all simple cases the Clebsch-Gordan colour factor is the matrix
element:
\begin{equation}
C_{F,r}(b,c;a_1,\ldots,a_r) = (T_{a_1} \ldots T_{a_r})_{bc} \; .
\end{equation}
The symmetrisation and calculation of colour factors will be carried
out separately when the vertex functions are reconstructed in a
Feynman diagram calculation.

The reduced vertex function has the structure:
{\bacol
\begin{eqnarray}
&& Y_{F,r}(\bm{p},\bm{q} ; 
\bm{k}_1,\mu_1;\ldots;\bm{k}_r,\mu_r) = 
\sum_{n=1}^{n_r} \Gamma_n
f_n  \times
\nonumber \\
&& ~~~~~~~~~~~~~~~~~~
\exp \left( \: \frac{i}{2} \left(
\bm{p} \cdot \bm{x} + \bm{q} \cdot \bm{y} +
\bm{k}_1 \cdot \bm{v}^{n}_1 + \ldots +
\bm{k}_r \cdot \bm{v}^{n}_r \right) \right) \; .
\label{eqn_sy}
\end{eqnarray}
}
As we do not use explicit representations of the spin matrices, it is
important that each monomial retains the correct spin dependence
$\Gamma_n$.

\subsection{Realistic actions: the gluon sector}

A general gluonic action is 
\begin{equation}
S(\psi,U) = \sum_{\bm{x}} 
\sum_{\mathcal{P}} c_{\mathcal{P}} \re \tr \left[ 
P(\bm{x},\bm{x},U) \right]  \; ,
\label{eqn_gen_glue}
\end{equation}
built of Wilson loops $P$ defined by closed paths
$\mathcal{P}(\bm{x},\bm{x};\bm{s})$, each with coupling constant
$c_{\mathcal{P}}$. The perturbative action is
\begin{eqnarray}
S_G(A) & = & \sum_r \frac{g^r}{r!} 
\sum_{\bm{k}_1,\mu_1,a_1} \ldots \sum_{\bm{k}_r,\mu_r,a_r}
\tA^{a_1}_{\mu_1}(\bm{k}_1) \ldots 
\tA^{a_r}_{\mu_r}(\bm{k}_r) \times
\nonumber \\
& & ~~~~~
V_{G,r}(\bm{k}_1,\mu_1,a_1; \, \ldots ; \, \bm{k}_r,\mu_r,a_r) \; .
\end{eqnarray}
The Euclidean Feynman rule for the $r$-point gluon vertex function is
$(- g^r V_{G,r})$, and the vertex $V_{G,r}$ is
\cite{Luscher:1986wf}
\begin{eqnarray}
&&
V_{G,r}(\bm{k}_1,\mu_1,a_1; \, \ldots ; \bm{k}_r,\mu_r,a_r)
= 
\nonumber \\
&& ~~~~~~
\frac{1}{r!} \sum_{\sigma \in \mathcal{S}_r}
\sigma \cdot C_{G,r}(a_1,\ldots,a_r) ~
\sigma \cdot Y_{G,r}(\bm{k}_1,\mu_1; \, \ldots ; \bm{k}_r,\mu_r) \; ,
\label{eqn_symmetrise}
\end{eqnarray}
The reduced vertex $Y_{G,r} = \sum_{\mathcal{P}} c_{\mathcal{P}}
Y_{G,r}^{\mathcal{P}}$ is the sum of contributions from paths
$\mathcal{P}$. As before, the $(r!)$ factor normalises the
symmetrisation.  $Y^{\mathcal{P}}_{G,r}$ can be expanded as
\begin{equation}
Y^{\mathcal{P}}_{G,r}(\bm{k}_1,\mu_1;\ldots;\bm{k}_r,\mu_r) = 
\sum_{n=1}^{n_r}
f_n  \; \exp \left( \: \frac{i}{2} \left( 
\bm{k}_1 \cdot \bm{v}^{n}_1 + \ldots +
\bm{k}_r \cdot \bm{v}^{n}_r \right) \right) \; .
\label{eqn_gy}
\end{equation}
In most cases we expect the lattice action to be real. For every
monomial in Eq.~(\ref{eqn_gy}), then, there must be a corresponding
term
\begin{equation}
(-1)^r f_n^* \exp - \left(\frac{i}{2}\sum_i \bm{k}_i \cdot \bm{v}_i^n
\right) \; .
\label{eqn_conj_ent}
\end{equation}
We can therefore speed up the evaluation of the Feynman rules by
removing the latter term, and replacing the exponentiation in
Eq.~(\ref{eqn_gy}) with ``$\cos$'' for $r$ even, and with ``$i \sin$''
for $r$ odd. Clearly we must identify to which terms this has been
applied.  This can either be done by recognising conjugate contours in
the action (e.g. $S = \half \tr [P + P^\dagger ]$) and expanding only
one, or by attaching a flag to each monomial to signal the reduction
(as discussed in Section~\ref{sec_python}).

If, in addition to the reality, the action has the form
Eq.~(\ref{eqn_gen_glue}) with a single trace in the fundamental
representation, the colour factors are
\begin{equation}
C_{G,r}(a_1,\ldots,a_r) = \half \left[ \tr \; ( T_{a_1} \ldots T_{a_r}) + 
(-1)^r \tr \; ( T_{a_r} \ldots T_{a_1}) \right] \; .
\label{cg_notwist}
\end{equation}
When symmetrising, a lot of the terms have similar Clebsch-Gordan
factors:
\begin{equation}
\sigma \cdot C_{G,r} = \chi_r(\sigma) \; 
C_{G,r}~~~~ \mathrm{where} ~~
\chi_r(\sigma) = \left\{ 
\begin{array}{l}
1~~~\mbox{for~~} \sigma \mbox{~~a cyclic permutation}, \\
(-1)^r~~~\mbox{for~~} \sigma \mbox{~~the inversion.}
\end{array}
\right.
\label{eqn_clebsch_sym}
\end{equation}
We can therefore partly symmetrise the vertex over $\mathcal{Z}_r$
(the subgroup of cyclic permutations and inversion) at the expansion
stage. The $\chi_r(\sigma)$ go into the amplitudes of the new terms
coming from the partial symmetrisation:
\begin{eqnarray}
V_{G,r}(\bm{k}_1,\mu_1,a_1; \, ... ; \,\bm{k}_r,\mu_r,a_r)
& = & \sum_{\sigma \in \mathcal{S}_r/\mathcal{Z}_r}
\sigma \cdot C_{G,r}(a_1,... ,a_r) \times
\nonumber \\
&& ~~~~~~ 
\sigma \cdot Y_{G,r}^\prime (\bm{k}_1,\mu_1;\, ... ;\,\bm{k}_r,\mu_r) \; ,
\nonumber \\
Y_{G,r}^\prime & = & 
\sum_{\stackrel{\mathcal{P}}{\sigma \in \mathcal{Z}_r}} \; 
c_{\mathcal{P}}
\chi_r(\sigma) \: \sigma \cdot Y^{\mathcal{P}}_{G,r} \; .
\label{v_notwist}
\end{eqnarray}
The advantage of doing this is that many of the extra monomials are
equivalent, and we can therefore cut down significantly the number of
exponentiation operations required to construct $V_{G,r}$. The number
of remaining symmetrisation steps (to be carried out in the Feynman
diagram code) is the number of cosets in $\mathcal{S}_r/\mathcal{Z}_r$
(one for $r \le 3$, three for $r=4$ etc.).

\subsection{Diagram differentiation}

There are many cases where Feynman diagrams need to be differentiated
with respect to one or more momenta. Whilst this can be done
numerically using an appropriately local difference operator, this can
lead to numerical instabilities.

It is clear from Eq.~(\ref{eqn_y}), that we can easily construct the
differentiated Feynman vertex. Let the momentum component we wish to
differentiate with respect to be $q_\nu$. We first construct a rank
$r$ object $\bm{\tau} = [\tau_1, \ldots, \tau_r]$ which represents the
proportion of momentum $\bm{q}$ in each leg of the Feynman diagram.
Momentum conservation dictates $\sum_i \tau_i = 0$.  For instance, for
a gluon 3-point function with incoming momenta
$(\bm{p},-\bm{p}+2\bm{q},-2\bm{q})$, we would have $\bm{\tau} =
[0,2,-2]$. The differentiated vertex is
\begin{eqnarray}
\frac{d}{dq_\nu} 
Y^{\mathcal{L}}_r(\bm{k}_1,\mu_1 ; \ldots ; \bm{k}_r,\mu_r) = 
& \sum_{n=1}^{n_r} & \
\frac{if_n}{2} \left(\tau_1 v^n_{1;\nu} + \ldots + \tau_r v^n_{r;\nu} 
\right) \times
\nonumber \\
&& ~~~~
\exp \frac{i}{2} \left(
\bm{k}_1 \cdot \bm{v}^n_1 + \ldots + \bm{k}_r \cdot \bm{v}^n_r \right)
\label{eqn_diff_y}
\end{eqnarray}
and so on for higher derivatives. We may therefore simultaneously
calculate as many differentials as we need for the cost of one
exponentiation. If this momentum expansion is placed into an
appropriate data structure with overloaded operations, it is easy to
create the Taylor series for a Feynman diagram by multiplying the
vertex factors together. For examples of such codes, see
\cite{ref_autodiff}.

It may also be necessary to differentiate diagrams with respect to
parameters in the action, which may be present in different multiples
in the amplitude of the monomials. Depending on the situation, we can
separately expand parts of the action containing different powers of
the parameter. Alternatively, we can use a single expansion and append
a label to each monomial that records how many powers of the given
parameter are contained in the amplitude. This label is then used in
the parameter differentiation in the Feynman diagram code.

\subsection{Recursive path definitions}
\label{sec_recursion}

So far we have assumed that the paths in the action are constructed
from single links. This is, of course, always true but it is often
more compact to specify the action as built from composite objects,
such as smeared links, giving a separate prescription for the link
smearing. For instance, the smeared link might be defined as the gauge
covariant, weighted sum of the link and its adjoining staples:
\begin{equation}
W_\mu(\bm{x}) = c_0 U_\mu(\bm{x}) + c_1 
\sum_{\stackrel{\pm \nu}{|\nu| \not= \mu} } 
U_\nu(\bm{x}) U_\mu(\bm{x} + \bm{e}_\nu) U^\dagger_\nu(\bm{x} + \bm{e}_\mu)
\; ,
\label{eqn_smear}
\end{equation}
where the coefficients $c_{0,1}$ define the smearing method and are
chosen to optimise certain aspects of the Monte Carlo simulation. The
smearing may also be defined recursively, with smeared links
inserted into additional smearing recipes. Examples include the
``HISQ'' improved staggered fermions discussed in
\cite{Follana:2004sz},
where the links are first ``FAT7'', then ``ASQ'' smeared. Rather than
multiplying out the paths in the action to give a large sum of
tangled paths built from links, it is more convenient to reflect the
nested improvement structure in the expansion algorithm itself.

This is done by first defining the mapping $U \to W$ as a sum of paths
as in Eq.~(\ref{eqn_path_def}).  For instance, Eq.~(\ref{eqn_smear})
is represented by path $\left\{ \bm{x},\bm{x}+\bm{e}_\mu;[\mu]
\right\}$ with coupling $c_0$ plus
$\left\{\bm{x},\bm{x}+\bm{e}_\mu;[\nu,\mu,-\nu] \right\}$,
$\left\{\bm{x},\bm{x}+\bm{e}_\mu;[-\nu,\mu,\nu] \right\}$ with
coupling $c_1$.  Call the smearing definitions $\mathcal{W}$.  An
action is then compactly specified by a path $\mathcal{P}$ of
composite objects.  These are in turn are defined by an ordered list
$[\mathcal{W}_1, \mathcal{W}_2, \ldots, \mathcal{W}_n]$ representing
each step of nested improvement.  The full field is then defined by
the recursion
\begin{equation}
\begin{array}{lll}
\mathcal{W}_1 & : U & = W_1(U) \; ,
\\
\mathcal{W}_2 & : W_1(U) & = W_2(W_1(U)) \; ,
\\
\mathcal{W}_3 & : W_2(U) & = W_3(W_2(U)) \; ,
\\
& & \vdots 
\\
\mathcal{W}_i & : W_{i-1}(U) & = W_i(W_{i-1}(U)) \; ,
\\
& & \vdots 
\\
\mathcal{P} & : W_n(U) & = P(W_n(U)) \; .
\end{array}
\label{eqn_recursion}
\end{equation}
\section{An expansion algorithm}
\label{sec_exp_wil}

In this section we present a new algorithm for carrying out the Taylor
expansion of lattice actions in a manner suited to computer
implementation.

We start by defining an object that represents a single term in the
Taylor expansion in Eq.~(\ref{eqn_y}). We call this an ``entity'' $E$,
and it is an ordered list:
\begin{equation}
E = \left( \mu_1,\ldots,\mu_r ; \bm{x},\bm{y} ; 
\bm{v}_1 , \ldots , \bm{v}_r ; f \right) \; .
\label{eqn_entity_defn}
\end{equation}
The order of the entity is $r$. For instance, a single link comes from
a path $\mathcal{L} = \{ \bm{0},\bm{e}_\mu ; [ \mu ] \}$, and the
$r^{\mathrm{th}}$ term in its expansion in Eq.~(\ref{eqn_link_exp}) is
represented as
\begin{equation}
E_r = 
( \underbrace{\mu,\ldots,\mu}_{r \mathrm{~terms}} \; ; 
\bm{0}, 2\bm{e}_\mu \; ; 
\underbrace{\bm{e}_\mu,\ldots,\bm{e}_\mu}_{r \mathrm{~terms}} \; ; 1 
) \;.
\end{equation}
Note that in units of $\frac{a}{2}$ the endpoint is $2 \bm{e}_\mu$ and
the midpoint $\bm{e}_\mu$. The reduced vertex function is a set of
all the entities of all orders in the expansion, and we call this a
``field'' $F = \left\{ E \right\}$.

In practise we build a Wilson line by concatenating smaller paths (the
smallest being the link). We therefore define the multiplication of
two fields so as to give the Taylor expansion of the resulting, longer
contour. The product is therefore the ordered product of each entity
from the first contour with every entity from the second:
\begin{eqnarray}
F(\mathcal{L}_1 \mathbin{*} \mathcal{L}_2) & = & 
F(\mathcal{L}_1) \mathbin{*} F(\mathcal{L}_2) \;,
\nonumber \\
& = & \left\{ E^i \mathbin{*} E^j ~~ \forall ~~ 
E^i \in F(\mathcal{L}_1),E^j \in F(\mathcal{L}_2) \right\} \; .
\end{eqnarray}
To keep gauge covariance, entity $E^j$ must be translated to start at
the end point of $E^i$. The order of the product entity is the sum of
those of the constituents. Further details are given in
Appendix~\ref{app_ent_alg}.

Addition of fields should represent the expansion of a sum of gluonic
paths. We therefore simply combine the lists of entities:
\begin{equation}
F(\mathcal{L}_1 + \mathcal{L}_2) = F(\mathcal{L}_1) +
F(\mathcal{L}_2) \;.
\end{equation}
In general, $F$ will be a redundant representation of the polynomial
containing two or more equivalent entities.  As each entity represents
a monomial which requires a computationally expensive exponentiation,
we construct a compression operation which compares all pairs of
entities in $F$ and combines them if they are equivalent:
\begin{eqnarray}
[E^i,E^j] & \longrightarrow & [E] ~~ \mathrm{if} ~~ E^i \equiv E^j
\nonumber \\
\mathrm{where} ~~ E & =  & \left(\mu^i_1 , \dots , \mu^i_r \; ;
\bm{x}^i,\bm{y}^i ; \bm{v}^i_1,\ldots,\bm{v}^i_r \; ; f^i+f^j \right)
\;.
\label{eqn_compression}
\end{eqnarray}
Assuming we have a translationally equivalent theory, entities need
only be equivalent up to translation by a constant vector. For
details, see Eq.~(\ref{eqn_xltn_inv}).

\section{A practical implementation}
\label{sec_python}

In this section we describe an implementation of the algorithm in a
programming language called \python. The \python\ interpreter is
freely available for a wide range of computational platforms at
\cite{ref_python}.
The complexity of some physical actions (notably high order NRQCD)
require the implementation to be CPU and memory efficient. This can be
achieved in \python\ with appropriate programming techniques, and
without sacrificing the object orientation and superior list handling
features of the language.

As a first step, we choose a maximum order for the Taylor expansion.
Any entity of higher order is discarded. The entities and all
sub-lists (including $D$-vectors) are encoded as ``tuples'', which are
immutable list objects to which the native hash function can be
applied.

To minimise the size of the dictionary, it is important that the field
does not contain entities that are equivalent. We choose data
structures for the entity and field specifically to prevent this.
Firstly, we exploit translational invariance of the action to arrange
that all paths start at the origin ($\bm{x}=\bm{0}$). Entity
equivalence in Eq.~(\ref{eqn_xltn_inv}) then follows from an item by
item tuple comparison.

This comparison is most efficiently done using a hash table (i.e. an
associative array).  The \python\ ``dictionary'' is a native
implementation of this, where a ``key'' indexes an ``object''. In our
implementation, each dictionary entry represents a monomial in the
reduced vertex function.  The key is a list of all information in the
entity bar the amplitude, which becomes the object indexed by this
key. Searching for equivalent entities now amounts to enquiring if the
key already exists in the dictionary. As all key entries are integer,
we do not need to worry about machine precision issues. We can
significantly speed up the hashing by omitting the now redundant
$\bm{x}=\bm{0}$ from the entity data structure. Independently, a
moderate performance gain can come from archiving the hash values for
each entity.

On a technical note, we find a slight speed-up associated with
implementing the field as a two-level dictionary. The upper level keys
are tuples of Lorentz indices. These each index a lower level
dictionary of all entities with the appropriate Lorentz structure.

If, on inserting a new entity into a field structure, an equivalent
entity exists, rather than adding the new item to the field its
amplitude is merely combined with that of the existing entity.  If the
new amplitude $f^i+f^j = 0$, the entry is removed.  This test is
robust for integer arithmetic. Otherwise, the absolute value is
compared to some tolerance, e.g.  $10^{-8}$. We may worry about
rounding errors: near $\bm{k}_i=0$ the reduced vertex monomials are
all adding in phase, and the deleted amplitudes may add to give a
significant contribution. We therefore use a tolerance smaller than is
finally required for the path expansion. As a last step only, we apply
the final tolerance cut. The numerical values of the intermediate and
final tolerances are found by trial and error, looking for robustness
in the number of terms in the expansion. It is worth pointing out that
\python\ carries out all floating point arithmetic in double
precision.

In the gluonic case of closed, traced contours the endpoint $\bm{y}$
is physically irrelevant. By ignoring it, we can identify more
entities that are equivalent and further reduce the dictionary size.
To find these, for each entity we impose $\bm{x} = \bm{y} = \bm{0}$
whilst translating all the $\bm{v}$'s such that $\bm{v}_1 =
\bm{e}_{\mu_1}$. The field dictionary is then rehashed, looking for
newly equivalent terms. Of course, care must be taken only to do this
as a final step, and not to then multiply such contours together.

As discussed in Section~\ref{sec_recursion}, a recursive action
definition is more compact and less error-prone. Each smearing
definition (such as ``APE'' or ``ASQ'') is predefined, and labelled
by a unique string. Each path in the action is specified by three
items: the coupling for this path, a list of signed directions and a
list of link improvement method names.  A full action (gluonic or
fermionic) is specified by a list of such path specifications. For
each path, the expansion routine is executed recursively to implement
Eq.~(\ref{eqn_recursion}), converting it into a field that is fed to
the next level of the nested link improvement.  Finally, these fields
can be manipulated or combined, before being output.

In some cases we need to combine complex conjugate monomials as per
Eq.~(\ref{eqn_conj_ent}). This is most easily done by noting that $E =
2 E_{\mathrm{real}} - E^*$. We loop over all unprocessed entities $E
\in F$, inserting $-E^*$ into $F$ and marking $E$ as now processed
(and doubling its amplitude). This marking can be done by adding a
Boolean flag to the entity data structure in
Appendix~\ref{app_ent_alg}.

Once we have the monomials, we have a choice of output format
dictated by whether the output is to be read into the Feynman diagram
evaluation code at compile time or run time.  The former has the
advantage that the compiler may be able to optimise the construction
of the vertex functions; the disadvantage is that the size of the
vertex functions may lead to code that is too long for the compiler to
handle. Reading in the data at run time avoids this, but may in
principle lead to slower code. In either case the \python\ code can be
easily adapted to produce the correct output format.

As an example, we describe the run time case. The output consists of a
single ASCII file for each order of the perturbative expansion. Each
file contains multiple entries, which could be a single line, or
multiple lines for clarity. The entry contains the information in a
single entity as whitespace separated values. For later storage in the
Feynman diagram code it is convenient if the Lorentz indices are
represented as a single integer in base D: $n(\bm{\mu}) =
\sum_{i=1}^r (\mu_i - 1) D^{i-1}$. It is also useful if the entries
for given $\bm{\mu}$ are consecutively numbered (although the order
does not matter). The file is terminated by a blank entry with
$n(\bm{\mu}) = -1$.

In the Feynman diagram code a set of arrays should be defined to hold
the vertex function data. For languages without allocatable arrays, we
can arrange for the \python\ to write a set of compile time header
files that create arrays of the correct dimensions for a given set of
vertex functions.

We use a set of Fortran90 modules to read in the data files, which can
serve as a template for other languages.

\section{Conclusions}
\label{sec_conc}

Simulation of Symanzik and radiatively improved lattice field theory
actions has become very popular in recent years. Associated
renormalisation factors (and, indeed, the radiative improvement
itself) can be systematically calculated using lattice perturbation
theory. The complicated nature of the improved actions and operators
has, however, contributed to a backlog in this perturbative
renormalisation programme.

Having a flexible method for generating the Feynman rules
automatically is crucial to overcoming this backlog, and permitting a
greater range of renormalisation factors to be calculated. This paper
provides just such a method, that is well suited to expanding all
sectors of lattice QCD: gluons, relativistic fermions and heavy
quarks. In addition to the Taylor expansion algorithm, an efficient
implementation in the \python\ programming language is described,
exploiting useful features of this language.

Particular strengths of this algorithm include coping with arbitrary
spin and colour trace structures in the action, allowing a nested
definition of link improvement and an intuitive way of defining the
action to be expanded.

The code is also very flexible, and can be adapted to deal with most
wrinkles met in perturbative expansion. The first of these is tadpole
improvement. Tadpole (or mean field) improvement aims to speed the
convergence of perturbation theory through dividing each link in the
action by a factor $u$
\cite{Lepage:1992xa}.
We can use perturbation theory to calculate $u = 1 + d_1 \alpha_s +
\dots$ as an expansion in the coupling, and treat the quantum effects
as radiative counterterms in the action with couplings $n_t d_i
(\alpha_s)^i$ (plus combinatoric factors), where $n_t$ was the number
of factors of links in the path. We can expand the action as before,
but must now do a separate expansion of the radiative counterterms.
Consider building an action from links smeared as in
Eq.~(\ref{eqn_smear}). When we expand $W_\mu(\bm{x}) W_\mu(\bm{x} +
\bm{e}_\mu)$ we will get a link combination
$[\nu,\mu,-\nu,\nu,\mu,\nu]$, and the question is whether such a term
should be given a tadpole improvement factor of $u^{-4}$ or $u^{-6}$.
The former is more in keeping with the philosophy of mean field
improvement, but in a simulation the latter is more convenient. The
procedure, of course, is that the perturbative action should follow what
is used in simulation. Either convention can be followed by assigning
to each entity knowledge of the full length of the path from which it
is derived. In the latter case of no cancellation, these lengths
simply add on entity multiplication. In the former case, some
directional knowledge must be maintained to allow factors of $u$ to
cancel. Terms with different numbers of tadpole factors should be
grouped separately; for this reason $n_t$ should be included in the
key in the \python\ implementation of the entities and fields.

This adaptability also makes the expansion algorithm described here
useful in, for instance, chiral perturbation theory 
\cite{Hippel:prog}
or the double expansion needed for stochastic perturbation theory
\cite{DiRenzo:2000ua,DiRenzo:2004ge}.

By describing and making available these algorithms and tools, we hope
that lattice field theory calculations can reach a point where the
choice of lattice action is not constrained by the availability (or
not) of renormalisation constants.

\section*{Acknowledgments}

We are pleased to acknowledge useful discussions with I.T. Drummond
and Q. Mason. A.H. is grateful to the U.K. Royal Society for support.



\appendix

\section{Spin algebra}
\label{app_spin_alg}

Each Wilson line in an action has an associated spin matrix. Where
this is not uniformly unity, we must keep track of which spin matrix
applies to each term in the reduced vertex. We do this by adding a
single integer label to the entity list, $\tau$.  For Dirac gamma
matrices $0 \le \tau \le 15$, whilst for Pauli sigma matrices $0 \le
\tau \le 3$. By convention, $\tau=0$ is the identity.

When two spin factors are multiplied, the product is proportional to a
single element of the spin algebra:
\begin{equation}
\tau^i \tau^j = \half \left[\tau^i,\tau^j \right] + 
\half \left\{\tau^i,\tau^j \right\} = 
\varepsilon_k \tau^k ~~ \mathrm{for~some} ~~
\varepsilon_k \in \mathbb{R}
\label{eqn_spin_mult}
\end{equation}
(with no sum over $k$). This reduction can be easily encoded in the
\python\ vertex generation code through a small dictionary where each
key $(\tau^i,\tau^j)$ indexes a list: $[\tau^k,\varepsilon_k]$.

\section{Pattern lists}
\label{app_patterns}

A Wilson line may be composed of a number of parts to which separate
colour traces may have been applied. This will affect the value of the
associated Clebsch-Gordan coefficient. Physical examples include the
traceless field strength operator, $U_{\mu \nu} - \tr U_{\mu \nu}$,
and the adjoint Yang--Mills action, $\tr U_{\mu \nu} \tr U_{\mu
  \nu}^\dagger$. In the former case, for instance, second order
monomials either have colour structure of the form $(T_a T_b)_{cd}$ or
$\tr (T_a T_b) \; \delta_{cd} = - \half \; \delta_{ab} \;
\delta_{cd}$.

As we do not calculate the Clebsch-Gordan coefficients in the vertex
generation program, we need a method for distinguishing whether given
gauge potentials within an entity are untraced, all traced together or
in separate colour traces. This distinction is also important in
ensuring we do not compress together entities with different colour
structures. We distinguish these cases by adding an extra entry to the
entity called a pattern list.

A pattern list of order $r$ is $\bm{\omega} = (\omega_1 , \ldots ,
\omega_r)$.  Each positive integer element $\omega_i$ is associated
with the corresponding factor of the gauge potential $A_i$. If $A_i$
has not been traced, $\omega_i = 0$. All gauge potentials with the
same value of the pattern component are understood to be contained in
a single colour trace. For instance, $A_1 A_3 \tr (A_2 A_4)$ would
have a pattern list $(0,1,0,1)$.

Applying a colour trace to an entity modifies only the pattern list
$\bm{\omega} \to \bm{\omega^\prime}$, with
\begin{equation}
\omega^\prime_i = \left\{
\begin{array}{l@{~~}l}
1 + \max(\omega_1, \ldots , \omega_r) \;, & \omega_i = 0 \;,
\\
\omega_i \;, & \omega_i \not= 0 \; .
\end{array}
\right.
\label{eqn_patt_tr}
\end{equation}
We stress that the actual value of $\omega_i$ has no meaning.  It is
therefore convenient to arrange at all stages that the first non-zero
element in the list is 1, the second 2 etc. Taking the example above,
$\tr [A_1 A_3 \tr (A_2 A_4)] = \tr (A_1 A_3) \tr (A_2 A_4)$ has a
relabelled pattern list $(1,2,1,2)$. Gauge invariance precludes
application of a trace to entities for which $\bm{x} \not= \bm{y}$
(i.e. to contours which are not closed).  The $SU(N)$ generators are
traceless, so when a colour trace applies to only one factor of the
gauge potential, we may delete the entity (such as when tracing an
entity of order $r=1$).

We define multiplication $\bm{\omega}^1 \mathbin{*} \bm{\omega}^2$ by:
\begin{equation}
(\omega^i_1 , \ldots , \omega^i_{r_i}) \mathbin{*}
(\omega^j_1 , \ldots , \omega^j_{r_j}) =
(\omega^i_1 , \ldots , \omega^i_{r_i},
\omega^\prime_1, \ldots , \omega^\prime_{r_j})
\label{eqn_mul_patt}
\end{equation}
where 
\begin{equation}
\omega^\prime_k = \left\{
\begin{array}{l@{~~}l}
0 \;, & \omega^j_k = 0 \;,
\\
\omega^j_k + \max (\omega^i_1 , \ldots , \omega^i_{r_i}) \;, & 
\omega^j_k \not= 0 \; .
\end{array}
\right.
\end{equation}
When all elements of $\bm{\omega}$ are not 1, the symmetries of
Eq.~(\ref{eqn_clebsch_sym}) are not present. The group of
symmetrisation operations carried out in the vertex generation code
must therefore be reduced from $\mathcal{Z}_r$, or it may be simpler
to postpone all symmetrisation until the Feynman diagram are evaluated
for specific momenta.

We note that pattern lists can also be used to label the taking of
real or imaginary parts in an action in a similar way, using positive
and negative entries to distinguish the two.

\section{Entity algebra}
\label{app_ent_alg}

Taking into account Appendices~\ref{app_spin_alg}
and~\ref{app_patterns}, an entity $E$ consists of
\begin{equation}
E = (\bm{\mu} ; \bm{x} , \bm{y} ;
\bm{v}_1 , \ldots , \bm{v}_r ; \bm{\omega} ; \tau ; f) \; .
\end{equation}
The colour trace pattern list $\bm{\omega}$ and spin index $\tau$ are
optional and need not be included in all situations. The start site
$\bm{x}$ may also be omitted (see Section~\ref{sec_python}).

The complex conjugate entity $E^*$ has amplitude $(-1)^r f^*$ and the
sign of all $D$-vectors reversed.

Multiplication by a scalar $p \in \mathbb{C}$ acts only on the amplitude:
\begin{equation}
p E = (\bm{\mu} ; \bm{x} , \bm{y} ; 
\bm{v}_1 , \ldots , \bm{v}_r ; \bm{\omega} ; \tau ; pf) \;.
\label{eqn_p_E}
\end{equation}
We translate an entity by $D$-vector $\bm{c} \in \Lambda$ using:
\begin{equation}
T_{\bm{c}} E = (\mu_1 , \ldots , \mu_r ; \bm{x} + \bm{c}, \bm{y} + \bm{c};
\bm{v}_1 + \bm{c}, \ldots , \bm{v}_r + \bm{c}; \bm{\omega} ; \tau ; f)
\label{eqn_Tc_E}
\end{equation}
Two entities are said to be equivalent if the lists can be rendered
identical under a translation and rescaling:
\begin{equation}
E^i \equiv E^j ~~~~ \mathrm{iff} ~~~~ \exists ~~ \bm{c} \in \Lambda, ~~
p \in \mathbb{C} ~~ \mathrm{s.t.} ~~
T_{\bm{c}} E^i = p E^j \; .
\label{eqn_xltn_inv}
\end{equation}
Non-commutative multiplication of two entities is defined by:
\begin{eqnarray}
E^\prime = E^i \mathbin{*} E^j & = &
( \bm{\mu}^\prime  ;
\bm{x}^i,\bm{y}^j + \bm{C} ; 
\nonumber \\
&& \bm{v}_1^i , \ldots , \bm{v}_{r_i}^i ,
\bm{v}_1^j + \bm{C}, \ldots , \bm{v}_{r_j}^j + \bm{C} ; 
\bm{\omega}^1 \mathbin{*} \bm{\omega}^2 ; \tau^k ;
f^\prime ) \; ,
\end{eqnarray}
\textit{i.e.} the path from which the second entity was derived is
first translated by a vector $\bm{C} = \bm{y}^i - \bm{x}^j$, to start
where the first finished. The resulting entity is of order $r = r_i +
r_j$. The Lorentz list $\bm{\mu}^\prime$ is the concatenation of lists
$\bm{\mu}^i + \bm{\mu}^j$. The spin indices yield $\tau^i \tau^j
= \varepsilon_k \tau^k$ as per Eq.~(\ref{eqn_spin_mult}). Note the
amplitude $f^\prime = \varepsilon_k \; {}^{r}C_{r_i} \; f^i \; f^j$
contains a combinatoric factor arising from having separated out the
$(r!)$ Taylor expansion factors from the amplitude in
Eq.~(\ref{eqn_tayl_exp}).

The action of the permutation operator
$\sigma=(\sigma_1,\sigma_2,\ldots,\sigma_r)$ on a list $\bm{l}$ yields
$(l_{\sigma_1}, \ldots , l_{\sigma_r})$. We can apply it to entities
which are closed $\bm{x} = \bm{y}$, simply traced ($\omega_i = 1 ~~
\forall ~~ i$) and where the real part has been taken (to ensure
Eq.~(\ref{cg_notwist}) holds)
\begin{equation}
\sigma \cdot E = 
( \sigma \cdot \bm{\mu} ; \bm{x},\bm{x} ; 
\bm{v}_{\sigma_1} , \ldots , \bm{v}_{\sigma_r} ; 
\sigma \cdot \bm{\omega} ; \tau ; 
\chi_r(\sigma) f ) \;,
\label{eqn_sigma_E}
\end{equation}
noting in this case that $\sigma \cdot \bm{\omega} \equiv
\bm{\omega}$. Eq.~(\ref{eqn_clebsch_sym}) defines $\chi_r(\sigma)$.

By extension Eqs.~(\ref{eqn_p_E},~\ref{eqn_Tc_E},~\ref{eqn_sigma_E})
and the colour trace are applied to a field $F$ by operating on each
of its constituent entities $E \in F$.








\end{document}